\def\chandra{{\it Chandra\/}}

\def\heao1{{\it HEAO-1\/}}
\def\hst{{\it {\it HST}\/}}

\def\ltsima{$\; \buildrel < \over \sim \;$}
\def\simlt{\lower.5ex\hbox{\ltsima}}
\def\gtsima{$\; \buildrel > \over \sim \;$}
\def\simgt{\lower.5ex\hbox{\gtsima}}

\documentclass{aastex}
\usepackage{emulateapj5,times,mathptm}

\begin{document}


\title{The Chandra Deep Field North Survey. VIII. 
  X-ray Constraints on Spiral Galaxies from $0.4 < z < 1.5$  }


\author{A.E.~Hornschemeier,$^{1}$ ~
W.N.~Brandt,$^1$
D.M.~Alexander,$^1$
F.E.~Bauer,$^{1}$
G.P.~Garmire,$^1$
D.P.~Schneider,$^1$
M.W.~Bautz,$^2$ and
G.~Chartas$^1$
}

\footnotetext[1]{Department of Astronomy \& Astrophysics, 525 Davey Laboratory, 
The Pennsylvania State University, University Park, PA 16802}

\footnotetext[2]{Massachusetts Institute of Technology, Center for Space Research, 
70 Vassar Street, Building 37, Cambridge, MA 02139}


\begin{abstract}

We present a statistical X-ray study of spiral galaxies in the Hubble 
Deep Field-North and its Flanking Fields using the Chandra Deep 
Field~North 1~Ms dataset.  We find that $\approx L^{*}$ galaxies with 
$0.4 \simlt z \simlt 1.5$ have ratios of X-ray to $B$-band luminosity 
similar to those in the local Universe, although the data indicate a likely
increase in this ratio by a factor of  
$\approx 2$--3.  We have also determined that 
typical spiral galaxies at $0.4 \simlt z \simlt 1.5$ should be 
detected in the 0.5--2 keV band in the flux range 
($3$--6$)\times10^{-18}$ erg~cm$^{-2}$~s$^{-1}$.  
\end{abstract}




\section{Introduction \label{intro}}

X-ray studies of fairly ``normal" galaxies, with high-energy emission
not obviously dominated by a luminous active galactic nucleus (AGN),
have recently been extended to cosmologically interesting distances in
the \chandra\ Deep Field (CDF) surveys, which have now reached 1~Ms of
exposure (CDF-N: Hornschemeier et~al. 2001, hereafter Paper~II; Brandt
et~al. 2001b, hereafter Paper~V; CDF-S: Tozzi et~al. 2001; P. Rosati
et~al., in prep.).
  Galaxies with  $0.1 \simlt z  \simlt 1.0$ are detected in appreciable
numbers at 0.5--2~keV fluxes below $1 \times 10^{-15}$
erg~cm$^{-2}$~s$^{-1}$ (e.g., Paper~II); the CDF-N survey goes almost two orders of
magnitude fainter, detecting significant numbers of normal galaxies
among the population of X-ray sources making the diffuse X-ray
Background (XRB; Paper~II; A.J. Barger et al.,  in prep.).  These
normal galaxies contribute as much as 5--10\% of the XRB flux in the
0.5--2~keV band.  The bulk of the energy density of the XRB is
certainly explained by AGN,  
but the investigation of the ``typical" galaxy, whether
its X-ray emission is dominated by  a population of X-ray binaries, hot
interstellar gas, or even a low-luminosity AGN, is an equally important
function of deep X-ray surveys.   Normal galaxies are likely to be the
most numerous extragalactic X-ray sources in the Universe and are
expected to dominate the number counts at 0.5--2~keV fluxes of $\approx
1 \times 10^{-17}$--$1 \times 10^{-18}$ erg~cm$^{-2}$~s$^{-1}$ (Ptak
et~al. 2001).

The CDF-N has reached the depths necessary to detect individually many normal 
[$\log{({{L}_{\rm X}\over{{L}_{\rm B}}})} \approx -3$; ${L}_{\rm X}$ 
is from 0.5--2 keV] $L^{*}$ galaxies to $z\approx0.3$, corresponding to a 
look-back time of $\approx 4$~Gyr ($H_0=65$~km~s$^{-1}$ Mpc$^{-1}$,
$\Omega_{\rm M}=1/3$, and
$\Omega_{\Lambda}=2/3$
are adopted throughout this paper).
 Reaching larger look-back times 
presents  the exciting possibility of detecting the bulk X-ray response 
to the heightened star-formation rate at $z\approx1.5$--3 (e.g., 
Madau et al. 1996).  One thus expects the X-ray luminosity per unit 
$B$-band luminosity to be larger at $z\approx0.5$--1 in the past due to the increased 
energy output of X-ray binary populations at $z\approx1.5$--3; this X-ray
emission represents a ``fossil record" of past epochs of star formation 
(e.g., Ghosh \& White 2001; Ptak et al. 2001).  Therefore, measurements 
of the X-ray luminosities of typical galaxies can constrain models of 
X-ray binary production in galaxies.

While X-ray emission from individual galaxies is not easily detected at 
$z \approx 1$, it is possible to estimate the emission at their extremely
faint flux levels using statistical methods such as stacking, a
technique implemented successfully on the CDF-N survey data in several
previous studies.  These include the detection of X-ray emission from
the average $z\approx0.5$ bright ($I \simlt 22$) galaxy in the
Hubble Deep Field-North (\hbox{HDF-N}) described in Brandt et~al. (2001a, hereafter Paper~IV) and
a study of X-ray emission from $z=2$--4 Lyman break galaxies identified
in the \hbox{HDF-N} (Brandt et~al. 2001c, hereafter Paper~VII).
Encouraged by the success of these analyses, we extend here the study
of normal galaxies to the entire \hbox{HDF-N} plus \hst\ Flanking Fields region, 
now concentrating on galaxies at $z < 2$ to complement
the study of $z > 2$ galaxies performed in Paper~VII.  We focus on this
redshift range due to the extensive spectroscopic redshift coverage (Cohen et~al. 2000 
and references therein) and superb \hst\ imaging which has
allowed a comprehensive galaxy morphology study (van den Bergh, Cohen,
\& Crabbe 2001).  The CDF-N data provide extremely deep X-ray coverage
over this area (see Figure~7 of Paper~V for the exposure map of this
region); the point-source detection limits in this region of the CDF-N
survey in the 0.5--2~keV and 2--8~keV bands are $\approx 3 \times
10^{-17}$~erg~cm$^{-2}$~s$^{-1}$ and $\approx 2 \times
10^{-16}$~erg~cm$^{-2}$~s$^{-1}$, respectively.  

In this study, we place observational constraints on the evolution of 
the ratio of X-ray luminosity to $B$-band luminosity of ``normal" spiral galaxies 
up to $z\approx 1.5$; this ratio is an indicator of the current level 
of star formation in a galaxy (e.g., David, Jones, \& Forman 1992; Shapley et~al. 2001).
We also place constraints on the fraction of the diffuse XRB explained
by galaxies lingering just below the CDF-N detection threshold, and thus
the contribution to the XRB by normal galaxies.


\section{Galaxy Samples}

\subsection{Redshift and Photometric Data}

Spectroscopic redshifts  for the galaxies are drawn from the catalogs
of Cohen et al. (2000), Cohen (2001), and Dawson et al. (2001) in the
range $0.1 < z < 1.1$.  Spectroscopic redshift determination is
difficult in the range $1.0 \simlt z \simlt 2.0$ due to the absence of
strong features in the observed-frame optical band and the lack of the
Lyman break feature useful to identify higher redshift objects.
 We have therefore used the deep photometric redshift catalog of
Fern\'andez-Soto, Lanzetta, \& Yahil (1999) for the redshift interval
$0.5 < z < 2.0$, which allows some overlap in redshift space with the
spectroscopic catalogs for cross-checking.   The spectroscopic 
catalogs cover the entire HDF-N plus a substantial fraction of the
\hst\ Flanking Fields region, whereas the photometric catalog only
covers the HDF-N.  We shall refer to these two samples as the
``spectroscopic sample" and the ``photometric sample" throughout the
rest of this Letter.

For the spectroscopic sample,  the $I$-band magnitude was used to
filter the sources by optical luminosity, as this is best matched to
rest-frame $B$ over most of the redshift range under consideration
here. The $I$ magnitudes are those given in Barger et al. (1999) for
the Hawaii Flanking Fields area.  For the photometric sample, the
\hst\ F814W (hereafter  $I814$)  magnitudes of Fern\'andez-Soto et al. (1999) were used.

\subsection{Sample Definition}

We chose galaxies which had no X-ray detection within 4\farcs0 
in the 0.5--8~keV (full), 0.5--2~keV (soft) and 2--8~keV
(hard) bands down to  a {\sc wavdetect} (Freeman et al. 2002)
significance threshold of $1 \times 10^{-5}$ in the restricted ACIS
grade set of Paper IV.
 This low detection threshold ensures that our study does not include
sources with X-ray emission just below the formal detection limits of
Paper~V.

We have attempted to construct a sample of galaxies similar to spiral
galaxies in the local Universe.  To accomplish this, we have used the
morphological classes of van den Bergh et~al. (2001) for galaxies from
$0.2 < z < 1.1$ in the HDF-N and the \hst\ Flanking Fields.  To
simplify the morphological filtering, we have cast objects in the 
van den Bergh et~al. (2001) catalog into the following four classes:  (1)
``E/S0" and ``E",
 (2) ``merger",  (3) ``Sa"--``Sc", including proto-spirals and
 spiral/irregulars, and (4) ``Irr",  ``peculiar" and/or ``tadpole".  We
 then filtered the catalog to keep only classes (2) and (3).

Filtering the photometric sample is more difficult due to the faintness
of many of the sources and problems due to morphological evolution with
redshift.  We have used the spectral energy distribution (SED)
classifications of Fern\'andez-Soto et al. (2001) to exclude all
galaxies of type ``E".  Comparison of the source lists reveals that,
within the area covered by both,
 $\approx$~70\% of galaxies identified through the two 
methods are in common.

Since the evolution of X-ray properties with redshift is of interest,
we have made an effort to study objects with comparable optical
luminosities at different redshifts.   This is particularly important
due to the non-linear relationship between X-ray luminosity and $B$-band
luminosity for some types of spiral galaxies ($L_{\rm X} \propto L_{\rm B}^{1.5}$;
e.g. Fabbiano \& Shapley 2001). Using the value of $M_{*}$ in the
$g^{*}$-band as determined by Blanton et~al. (2001) for a large sample of
galaxies in the Sloan Digital Sky Survey, we determined the value of  
 $M_{*}$ in the $B$-band.  The Sloan filter $g^{*}$ is best matched to
$B$;  the resulting value of $M_{*}$ in the $B$ band is $-20.43 \pm 0.10$.  

\begin{figure}
\figurenum{1}
\includegraphics[scale=0.85,angle=0]{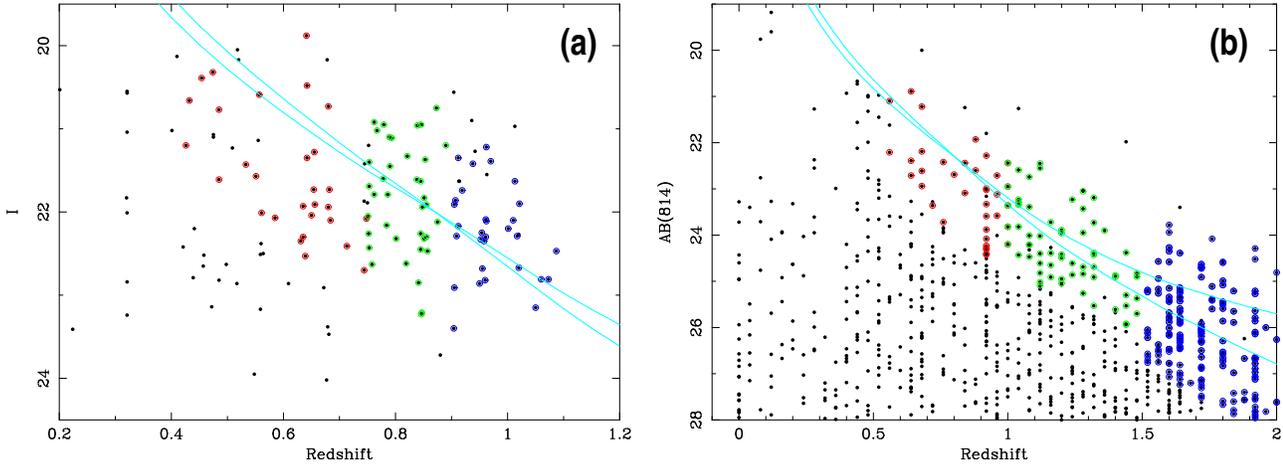}

\caption{Galaxies used in the stacking analyses. 
(a) The black filled circles are the spectroscopic
samples of Cohen et~al. (2000), Cohen (2001), and Dawson et~al. (2001) filtered
by morphology using van den Bergh et~al. (2001).  The
galaxies used in the stacking analyses are marked with open colored circles;
the color indicates the redshift bin, as described in 
Table~\ref{sample_table}. The blue curves give $I$ vs. $z$ 
for $M_{*}$ Sa and Sc galaxies (the lower curve at higher redshift is for the Sa galaxy).  
Galaxies without a colored circle were
not included in the stacking sample because they either were not within the range of optical
luminosity specified or because an X-ray detection was found within 4\farcs0.
(b) The black filled circles are the photometric redshift sample of Fern\'andez-Soto et~al. (1999), 
excluding the ``E"-type galaxies.
The blue curves give $I814$ vs. $z$ for $M_{*}$ Sa and Sc galaxies 
(the lower curve at higher redshift is for the Sa galaxy).
\label{sample_definition}}
\end{figure}

To ensure that our results are not sensitively dependent upon the 
galaxy SED used to determine the optical properties, we have 
used both the Sa and Sc galaxy SEDs of Poggianti et al. (1997) to calculate
$I$ and $I814$ vs. $z$ for an $M_{*}$ galaxy using the synthetic
photometry package SYNPHOT in IRAF.  These calculations are shown
in Figure~\ref{sample_definition}. Note the close similarity between the
Sa and Sc tracks; this is because the $I$ band corresponds 
to rest-frame $B$ in the middle of our redshift range.

Also plotted in Figure 1 are the 151 galaxies in the spectroscopic sample
with spiral or merger morphology having $z \leq 1.1$ and the 651 galaxies in the photometric 
sample with SED class other than ``E" having $0.5 \leq z \leq 2.0$.
These galaxies were filtered by optical flux to lie within 1.5 mags of the 
$M_{*}$ galaxy tracks discussed above; the galaxy samples constructed 
assuming Sa and Sc SEDs were identical (or nearly so) for all redshifts up to $z\approx 1.2$. 
Galaxies meeting the optical magnitude filter were then divided by redshift 
into several bins; these bins were constructed to ensure that 
there were $\simgt 30$ galaxies per bin.  The number of 
galaxies, median redshift, median look-back time and median optical magnitude 
for each bin are listed in Table~1.  In Figure~1, we mark all the objects in the
Sc SED sample with colors indicating the different redshift bins.

Table~1 also includes the number of galaxies rejected from 
each redshift bin due to the presence of an X-ray detection within 4\farcs0;
this exclusion radius ensures that our results will not be adversely
affected by the wings of the PSF of very bright X-ray sources.
These galaxies satisfied both the optical magnitude and morphology
filtering constraints and were rejected only due to X-ray detection.  
This exclusion criterion is very conservative, however, 
considering that our astrometry is accurate to $\approx 0$\farcs6
 in the area under consideration (see Paper~V).  To allow for the off-nuclear
nature of some of the X-ray sources found in normal galaxies (e.g., Paper IV),
we consider galaxies to be highly confident X-ray detections if the X-ray 
source is within 1\farcs5 of the galaxy's center.  This 
matching radius is also well matched to the Chandra PSF.\footnote{This radius
corresponds to the 90\% encircled-energy radius at $3^{\prime}$
off-axis for 0.5--2 keV and the 83\% encircled-energy radius for 0.5--8 keV.} 
We therefore also give the number of galaxies having an X-ray detection within 
1\farcs5 in Table~1.

\section{Stacking Procedure and Results }

The X-ray imaging data at each position were stacked in the same manner
as in Paper~VII, keeping the 30 pixels whose centers fall within an
aperture of radius 1\farcs5.  The detection significance 
in each band was assessed by performing 100,000 Monte-Carlo stacking
simulations using local background  regions as in Paper~VII.  A source
is considered to be significantly detected if the number of counts over
background exceeds that of 99.99\% of the simulations.  No single source 
in the stacking sample appeared to dominate the distribution, 
demonstrating the effectiveness of our selection criteria.

Stacking of the galaxies in the redshift bins described in Table~\ref{sample_table}
and Figure~\ref{sample_definition} resulted in significant detections
in the soft band for all of the redshift bins up to $z\approx
1.5$ (see Table~2).  We also stacked galaxies in the redshift bin 
$1.5 < z < 2.0$, but there was not a significant detection.
The results for the two different spiral galaxy SED
samples are nearly or exactly identical except for the detection in the
highest redshift bin ($1.0 < z < 1.5$). We adopt a $\Gamma=2.0$ power law for the calculation 
of X-ray fluxes and luminosities, assuming that these galaxies 
are similar to spiral galaxies in the local Universe and
have their X-ray emission dominated by X-ray binaries (e.g., Kim, Fabbiano, \& Trinchieri 1992).
While there were several cases of significant detections in the full band,  there were no
highly significant detections in the hard band.  Given the variation
of effective area and background rate with energy, the signal-to-noise
ratio for sources with the assumed spectrum is highest in the soft band
and lowest in the hard band, so this behavior is expected.
The flux level of the soft-band detections for the spectroscopic sample is
 (5--6)$\times10^{-18}$ erg~cm$^{-2}$~s$^{-1}$.   The corresponding
rest-frame 0.5--2~keV luminosities for the average galaxy 
are $\approx 1\times 10^{40}$ erg~s$^{-1}$ for the
lowest redshift bin and  $\approx 2\times 10^{40}$ erg~s$^{-1}$ for the
highest redshift bin.  For the photometric sample, the soft-band flux level of
the detections is (3--5)$\times10^{-18}$ erg~cm$^{-2}$~s$^{-1}$.
We also give fluxes for the less significant detections in the
full band for those redshift bins having highly significant
soft-band detections.

We have investigated the properties of the sources which were rejected from
the stacking samples due to an X-ray detection at or near the position
of the galaxy (the numbers of such sources are given in the last column
of Table~1).   There are only 15 distinct galaxies with X-ray
detections within 1\farcs5.  Of these 15
sources, three are broad-line AGN, which are clearly not normal
galaxies.  One object has a photometric redshift which differs 
significantly from its spectroscopic redshift.  Since the optical
properties of this object at its spectroscopic redshift place it
outside our sample boundaries, we have rejected it.  One object is 
very near another X-ray source which has been positively identified
with a narrow-line AGN.  Thus, there are a total of 10 ``normal"
galaxies positively identified with X-ray sources within this
sample, consituting a small minority of the galaxies under study here.
The worst case is in the lowest redshift bin
where 15\% of the galaxies had X-ray detections.

\begin{figure}
\figurenum{2}

\includegraphics[scale=0.80,angle=0]{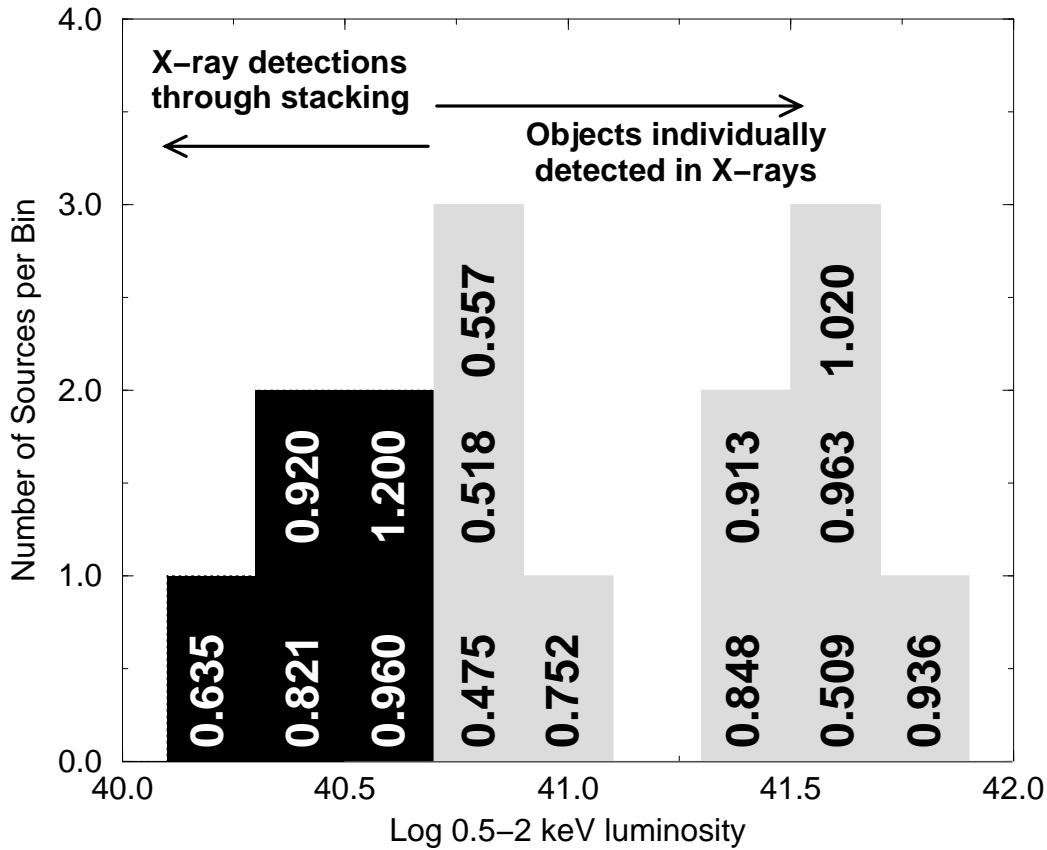}

\caption{
Histogram of X-ray luminosities obtained both from the stacking analysis (black)
and from individually detected galaxies (grey).  Source redshifts for
each bin are printed over the histogram points in that bin.
For the purpose of comparison, we treat each of the average stacking 
detections as single ``sources;" these are labeled with the median redshifts
of the sources which were stacked to yield the detection.  
 We compare these average stacked properties with the properties 
of the individually detected sources. The 10 X-ray 
detected sources are at higher X-ray luminosities and are not typical of the galaxies which
were stacked, however there are seeral objects at $z\approx0.5$ which appear to
be plausibly drawn from the same X-ray luminosity distribution as the stacked galaxies. 
}
\label{sbhistogram}
\end{figure}

 Figure~\ref{sbhistogram} shows a histogram of $L_{\rm X}$ values
 calculated for both the stacking samples and the individually X-ray
detected galaxies.  The individually X-ray detected galaxy set does
possibly contain some lower-luminosity AGN, including the AGN candidate
\hbox{CXOHDFN~J123643.9+621249} (see Papers II and IV).  With the exception
of the objects in the redshift bin with median $z=0.635$, Figure~\ref{sbhistogram} 
shows that typically the X-ray
luminosities of the individually detected objects are on average much
higher than those of the stacked galaxies; they are sufficiently more
luminous as to appear atypical of the normal galaxy population. For the
lowest redshift bin, it is plausible that our results are 
moderately biased by the exclusion of the X-ray detected sources.
In \S 4, we therefore also give results which include  
the individually X-ray detected objects in the sample average for
this lowest redshift bin.

 For additional comparison, we have considered the radio properties of
the individually detected galaxies and the stacked galaxies using
the catalogs of Richards et~al. (1998) and Richards (2000).  The
percentage of radio detections among the individually X-ray detected
galaxies is higher than that among the stacked galaxies ($\approx 27$\% vs.
$\approx2$--5\%). This possible difference between the two
populations is significant at the 93\% level as determined using the
Fisher exact probability test for two independent samples   
(see Siegel \& Castellan 1988).
 Due to the X-ray luminosity difference and
possible difference in radio properties, and to the fact
that these galaxies constitute a small minority of those under study,
we are confident we have not biased our determination of the properties
of the typical galaxy by omitting these X-ray detected objects from
further consideration.

\section{Discussion}

\begin{figure}
\figurenum{3}

\includegraphics[scale=0.85,angle=0]{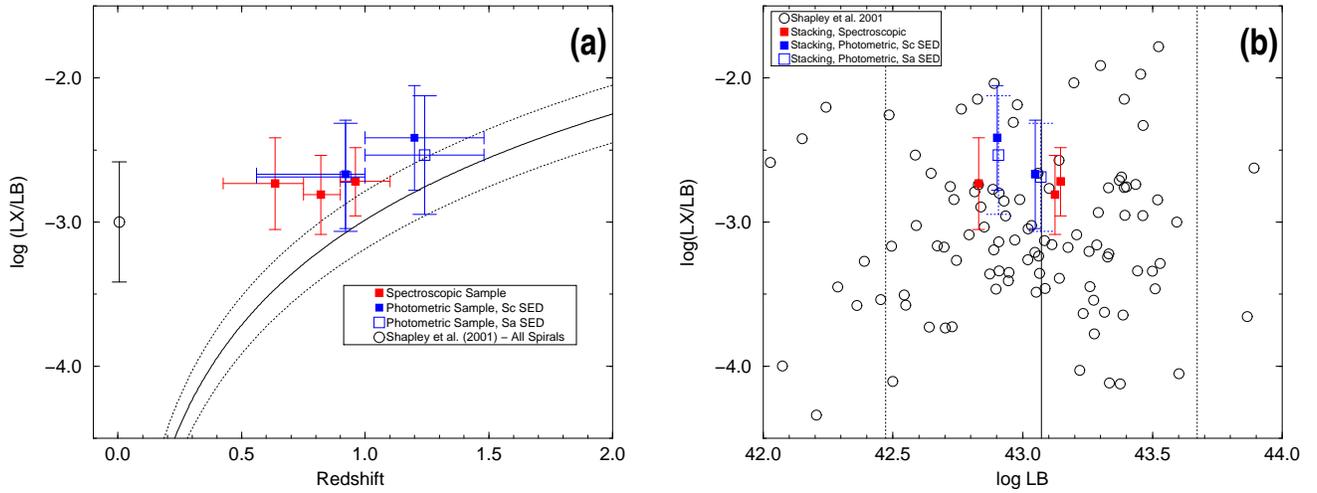}
\caption{ (a) $\log{({{L}_{\rm X}\over{{L}_{\rm B}}})}$ as a function of redshift
for the stacking samples.  The redshift error bars indicate the full extent of
the redshift bin; the data points are at the median redshift value for that bin.
The solid line indicates the 2$\sigma$ X-ray sensitivity limit
normalized by ${L}_{\rm B}^{*}$. The dashed lines above and below the solid line
indicate the effect of decreasing and increasing the optical luminosity by one magnitude, respectively.
Objects which have less X-ray luminosity per
unit $B$-band luminosity than this are not
expected to be detected in the current data.  
The error bar on the Shapley et al. (2001) data point indicates the dispersion of
values in this sample.
  (b)  $\log{({{L}_{\rm X}\over{{L}_{\rm B}}})}$ 
vs. ${L}_{\rm B}$ for both the Shapley et al. (2001) sample (open circles) and
the stacked detections presented here.
The error bars on $\log{({{L}_{\rm X}\over{{L}_{\rm B}}})}$ in both figures were 
calculated following the numerical method described in \S 1.7.3. of Lyons (1991).
The solid line in (b) indicates ${L}_{\rm B}^{*}$; again the dashed lines correspond to one
magnitude fainter and brighter than ${L}_{\rm B}^{*}$. 
\label{lxlb}
}

\end{figure}

In Figure~\ref{lxlb}a, we plot the X-ray-to-optical luminosity 
ratio $\log{({{L}_{\rm X}\over{{L}_{\rm B}}})}$ for each 
stacked detection, where ${L}_{\rm X}$ is calculated for the rest-frame 0.5--2~keV 
band. We have also plotted our approximate sensitivity limit in Figure~\ref{lxlb}a,
which is simply the corresponding $2 \sigma$ X-ray luminosity
detection limit achieved for a 30~Ms stacking analysis divided by ${L}_{B}^{*}$.
We do not expect to detect galaxies having less X-ray emission per unit $B$-band
luminosity than this value.  For comparison with the local Universe, we also 
plot the mean $\log{({{L}_{\rm X}\over{{L}_{\rm B}}})}$ for spiral galaxies 
of comparable $L_{\rm B}$ from the sample of Shapley et~al. (2001). 
This sample includes 234 spiral galaxies observed  
with {\it Einstein} and excludes AGN where 
the X-ray emission is clearly dominated by the nucleus.  
The galaxies in the Shapley et al. (2001) sample all have $z < 0.025$;
median redshift is $z\approx0.004$.

In Figure~\ref{lxlb}b, 
we plot $\log{({{L}_{\rm X}\over{{L}_{\rm B}}})}$ versus ${L}_{\rm B}$; 
the values up to $z \approx 1.0$ are consistent with what is
observed in the Shapley et~al. sample for objects with comparable ${L}_{\rm B}$, although they are
toward the high end of what is observed.  This is consistent with
Figure \ref{lxlb}a, which shows the average $\log{({{L}_{\rm X}\over{{L}_{\rm B}}})}$
 derived from stacking being somewhat higher than that for the Shapley et~al. galaxies of
comparable optical luminosity.  There is a slight increase (factor of 1.6) in the average
${{ {L}_{\rm X}\over{{L}_{\rm B}} }}$ from the local Universe to $z\approx0.6$.
For the highest redshift bin
($1.0 < z < 1.5$), the results become somewhat sensitive to the galaxy
SED one assumes for determining the optical properties.  We suspect
that an Sc galaxy SED is more appropriate at this epoch due to the
higher prevalence of star formation.  Adopting this SED, we find that
${{{L}_{\rm X}\over{{L}_{\rm B}} }}$ has increased
somewhat more substantially ($\approx 3.8$ times) at $1.0 < z < 1.5$.

One may also constrain star-formation models using only the X-ray 
properties of these galaxies.  The average X-ray luminosity of the 
spiral galaxies in the Shapley et~al. (2001) sample having the same 
range of ${L}_{\rm B}$ as used in this study is 
$7.8\times10^{39}$~erg~s$^{-1}$ (converted to 0.5--2 keV).  The 
average galaxy in our stacking sample has a luminosity
$\approx 1.6$ times higher at $z\approx0.6$.  This increases 
to $\approx 3.4$ times higher at $z\approx1.0$.  The $z\approx0.6$ value
 is most likely affected by some bias due to the exclusion of 
legitimate normal galaxies in the lowest redshift bin (see \S 3).  If we include
these individually X-ray detected objects, then the average galaxy in our stacking sample has a luminosity
$\approx 3.3$ times higher at $z\approx0.6$, 
 consistent with the predictions of Ghosh \& White (2001) that the X-ray 
luminosity of the typical Sa-Sbc spiral galaxy should be $\approx 3.3$ 
times higher at $z\approx 0.5$.  Including the X-ray detected galaxies
does not significantly affect our results for the interval $0.75 < z < 0.90$ (the
difference is $\approx35$\%).  However, since the X-ray luminosities
of the X-ray detected objects with $z\approx1.0$ are substantially higher
(by an order of magnitude) than the X-ray stacking averages (see Figure~2),
it is not appropriate to include these objects in the calculation of the 
average X-ray luminosity. 
We thus find a smaller increase in the average X-ray luminosity of galaxies
at $z\approx1$ than the increase by a factor of $\approx 5.4$ predicted
by Ghosh \& White (2001).  We find that the average X-ray luminosities
of galaxies have not evolved upward by more than a factor of $\approx 3.4$
by $z\approx1.0$, regardless of exclusion or inclusion of X-ray 
detected objects.

 The range of average 0.5--2~keV fluxes for the spiral galaxies 
studied here is 
($3$--6$)\times10^{-18}$ erg~cm$^{-2}$~s$^{-1}$.  These X-ray fluxes
are consistent with independent predictions made by Ptak et al. (2001)
that galaxies of this type will be detected at 0.5--2 keV flux levels of
$\approx 6.6 \times 10^{-18}$ erg~cm$^{-2}$~s$^{-1}$ (converted from
their 2--10 keV prediction assuming a $\Gamma = 2.0$ power law).

Assuming a 0.5--2~keV XRB flux density of  
$6.95\times 10^{-12}$~erg~cm$^{-2}$~s$^{-1}$~deg$^{-2}$ 
(Garmire et al. 2001), we have identified $\approx 1$\% 
of the soft XRB as arising from spiral galaxies not yet 
individually detected in deep \chandra\ surveys.  Many of 
these objects should be sufficently bright to be detected 
with \chandra\ ACIS exposures of $\approx 5$~Ms, which 
should be achievable in the next several years of the mission.


\acknowledgments

We thank Alice Shapley for useful discussions and sharing data.
We gratefully acknowledge the financial support of
NASA grant NAS~8-38252 (GPG, PI),
NASA GSRP grant NGT5-50247 (AEH),
NSF CAREER award AST-9983783 (WNB, DMA, FEB), and 
NSF grant AST99-00703~(DPS). 
This work would not have been possible without the enormous efforts of the entire
\chandra\ team. 
%


\begin{deluxetable}{ccccccc}
\tabletypesize{\footnotesize}
\tablewidth{0pt}
\tablecaption {Stacking Redshift Bins \label{sample_table}}
\scriptsize
\tablehead{
\colhead{Redshift} & 
\colhead{} &
\colhead{Median} &
\colhead{Median} &
\colhead{Median} &
\colhead{ } &
\colhead{} \\
\colhead{Range} &
\colhead{SED$^{\rm a}$} &
\colhead{$z$} &
\colhead{$t_{\rm LB}$$^{\rm b}$} &
\colhead{$I$/$I814$$^{\rm c}$}  &
\colhead{$N^{\rm d}$ } &
\colhead{$N_{\rm REJ}$$^{\rm e}$}
}
\startdata
\multicolumn{7}{c}{SPECTROSCOPIC SAMPLE} \\
\hline 
  0.40--0.75 & Sc & 0.635 & 6.34  & 21.73 & 29 & 12/6 \\
  0.75--0.90 & Sc & 0.821 & 7.39  & 21.83 & 38 & 2/2 \\
  0.90--1.10 & Sc & 0.960 & 8.05  & 22.29 & 30 & 6/4 \\
\hline
\\
\multicolumn{7}{c}{PHOTOMETRIC SAMPLE} \\
\hline 
  0.50--1.00 & Sc & 0.920 & 7.87  & 22.99 & 37 & 9/4 \\
  1.00--1.50 & Sc & 1.200 & 8.97  & 24.28 & 64 & 6/2 \\
  1.00--1.50 & Sa & 1.240 & 9.09  & 24.69 & 80 & 8/2 \\  
\enddata

\tablenotetext{a}{Galaxy SED assumed for filtering by optical luminosity (see Figure~1).}
\tablenotetext{b}{Median look-back time in Gyr.}
\tablenotetext{c}{$I$ for the spectroscopic sample and
$I814$ for the photometric sample.}
\tablenotetext{d}{Number of galaxies satisfying the filter
criteria described in the text. This is the number of galaxies
in each stacking sample.}
\tablenotetext{e}{Number of galaxies rejected due to an
X-ray detection within 4\farcs0.
The second number indicates the number of these galaxies with X-ray detections within 1\farcs5.
Note that $N + N_{\rm REJ}$ is the total number of galaxies which would satisfy the optical
filtering criteria alone. }
\end{deluxetable}


\begin{deluxetable}{rrrrrrrrrcrrc}
\tabletypesize{\footnotesize}
\tablewidth{0pt}
\tablecaption {Stacking Results}
\tablehead{
\colhead{Redshift} &
\multicolumn{1}{c}{} &
\multicolumn{2}{c}{} &
\multicolumn{2}{c}{} &
\multicolumn{2}{c}{$t_{\rm Eff}^{\rm c}$ } &
\multicolumn{2}{c}{$<f_{\rm X}>^{\rm d}$ } & 
\multicolumn{2}{c}{$<L_{\rm X}>$ } &
\multicolumn{1}{c}{$<L_{\rm B}>^{\rm e}$ } \\ 
\colhead{Range} &
\multicolumn{1}{c}{SED} &
\multicolumn{2}{c}{Counts$^{\rm a}$ } &
\multicolumn{2}{c}{Significance$^{\rm b}$} &
\multicolumn{2}{c}{(Ms)} &
\multicolumn{2}{c}{(10$^{-17}$ erg~cm$^{-2}$~s$^{-1}$)} &
\multicolumn{2}{c}{(10$^{40}$ erg~s$^{-1}$)}  & 
\multicolumn{1}{c}{(10$^{43}$ erg~s$^{-1}$)} \\
\colhead{ } &
\colhead{ } &
\colhead{FB} &
\colhead{SB} &
\colhead{FB} &
\colhead{SB} &
\colhead{FB} &
\colhead{SB} &
\colhead{FB} &
\colhead{SB} &
\colhead{FB} &
\colhead{SB} \\
}
\startdata
\multicolumn{11}{c}{SPECTROSCOPIC SAMPLE} \\
\hline
0.40--0.75 & Sc & 49.6  & 31.7  &    99.99 &  $>99.99$ &  25.41 &  25.44 & 1.52 &  0.65 &2.94 & 1.26 & 0.67  \\
0.75--0.90 & Sc & 58.0  & 36.4  & $>99.99$ &  $>99.99$ &  33.05 &  33.09 & 1.36 &  0.57 &4.92 & 2.07 & 1.33 \\
0.90--1.10 & Sc & 54.4  & 25.9  & $>99.99$ &  $>99.99$ &  26.54 &  26.56 & 1.60 &  0.51 &8.44 & 2.68 & 1.40 \\
\hline
\\
\multicolumn{11}{c}{PHOTOMETRIC SAMPLE} \\
\hline
0.50--1.00 & Sc & 42.0 & 33.5  &    99.77 &  $>99.99$ &  34.51 &  34.54 & 0.95 &  0.51 & 4.51 & 2.40 & 1.12 \\
1.00--1.50 & Sc & 44.8 & 38.5  &    98.94 &  $>99.99$ &  59.67 &  59.73 & 0.58 &  0.34 & 5.34 & 3.06 & 0.79 \\
1.00--1.50 & Sa & 45.6 & 34.2  &    98.29 &  99.95    &  74.61 &  74.68 & 0.48 &  0.24 & 4.70 & 2.36 & 0.81 \\
\hline
\enddata

\tablenotetext{a}{Source counts are taken from the 30-pixel aperture
described in \S 3; background has been subtracted.  FB$=$0.5--8~keV, SB$=$0.5--2~keV}
\tablenotetext{b}{Significance levels are calculated by performing 100,000 stacking
simulations (see \S 3).
They reflect the fraction of the time one expects to obtain fewer than the observed number of
counts from randomly placed apertures in the absence of sources.}
\tablenotetext{c}{The effective exposure times are determined by stacking the 
appropriate band's exposure map in a manner similar to the stacking of the X-ray imaging data;
the average of the effective exposure times within the aperture is given.}
\tablenotetext{d}{Fluxes are calculated assuming a $\Gamma=2$ power-law model and are for the
rest frame. }
\tablenotetext{e}{For reference, ${L}_{B}^{*} = 1.09 \times 10^{43}$~erg~s$^{-1}$ (see \S 2.2).}

\end{deluxetable}

\clearpage

\end{document}